\documentclass[aps,twocolumn,showpacs,superscriptaddress,groupedaddress]{revtex4}  
\usepackage{graphicx}
\usepackage{subfigure}
\usepackage[breaklinks=true,colorlinks=true,linkcolor=blue,urlcolor=blue,citecolor=blue]{hyperref}
\usepackage{bm}        
\usepackage{amssymb,amsmath,amsopn,amsfonts}

\newcommand{\Sref}[1]{Sec.~\ref{#1}}
\newcommand{\sref}[1]{Sec.~\ref{#1}}
\newcommand{\fref}[1]{Fig.~\ref{#1}}
\def\eqref#1{(\ref{#1})}
\begin{document}

\title{Full reconstruction of a 14-qubit state within four hours}

\author{Zhibo~Hou}
\affiliation{Key Laboratory of Quantum Information,University of Science and Technology of China, CAS, Hefei 230026, P. R. China}
\affiliation{Synergetic Innovation Center of Quantum Information and Quantum Physics, University of Science and Technology of China, Hefei 230026, P. R. China}
\author{Han-Sen Zhong}
\affiliation{Key Laboratory of Quantum Information,University of Science and Technology of China, CAS, Hefei 230026, P. R. China}
\affiliation{Synergetic Innovation Center of Quantum Information and Quantum Physics, University of Science and Technology of China, Hefei 230026, P. R. China}
\author{Ye~Tian}
\affiliation{Key Laboratory of Quantum Information,University of Science and Technology of China, CAS, Hefei 230026, P. R. China}
\affiliation{Synergetic Innovation Center of Quantum Information and Quantum Physics, University of Science and Technology of China, Hefei 230026, P. R. China}
\author{Daoyi~Dong}
\affiliation{School of Engineering and Information Technology, University of New South Wales, Canberra, ACT 2600, Australia}
\author{Bo Qi}
\affiliation{Key Laboratory of Systems and Control, ISS, and National Center for Mathematics and Interdisciplinary Sciences, Academy of Mathematics and Systems Science, CAS, Beijing 100190, People's Republic of China}
\author{Li~Li}
\affiliation{Centre for Quantum Computation and Communication Technology and Centre for Quantum Dynamics, Griffith  University,  Brisbane,  Queensland  4111,  Australia}
\author{Yuanlong~Wang}
\affiliation{School of Engineering and Information Technology, University of New South Wales, Canberra, ACT 2600, Australia}
\author{Franco~Nori}
\affiliation{CEMS, RIKEN, Saitama 351-0198, Japan}
\affiliation{Physics Department, University of Michigan, Ann Arbor, Michigan 48109-1040, USA}
\author{Guo-Yong~Xiang}
\email{gyxiang@ustc.edu.cn}
\affiliation{Key Laboratory of Quantum Information,University of Science and Technology of China, CAS, Hefei 230026, P. R. China}
\affiliation{Synergetic Innovation Center of Quantum Information and Quantum Physics, University of Science and Technology of China, Hefei 230026, P. R. China}
\author{Chuang-Feng~Li}
\affiliation{Key Laboratory of Quantum Information,University of Science and Technology of China, CAS, Hefei 230026, P. R. China}
\affiliation{Synergetic Innovation Center of Quantum Information and Quantum Physics, University of Science and Technology of China, Hefei 230026, P. R. China}
\author{Guang-Can~Guo}
\affiliation{Key Laboratory of Quantum Information,University of Science and Technology of China, CAS, Hefei 230026, P. R. China}
\affiliation{Synergetic Innovation Center of Quantum Information and Quantum Physics, University of Science and Technology of China, Hefei 230026, P. R. China}

\vspace{10pt}
\begin{abstract}
Full quantum state tomography (FQST) plays a unique role in the estimation of the state of a quantum system without \emph{a priori} knowledge or assumptions. Unfortunately,  since FQST requires informationally (over)complete measurements, both the number of measurement bases and the computational complexity of data processing suffer an exponential growth with the size of the quantum system. A 14-qubit entangled state has already been experimentally prepared in an ion trap, and the data processing capability for FQST of a 14-qubit state seems to be far away from practical applications. In this paper, the computational capability of FQST is pushed forward to reconstruct a 14-qubit state with a run time of only 3.35 hours using the linear regression estimation (LRE) algorithm, even when informationally overcomplete Pauli measurements are employed. The computational complexity of the LRE algorithm is first reduced from $\sim 10^{19}$ to $\sim 10^{15}$ for a 14-qubit state, by dropping all the zero elements, and its computational efficiency is further sped up by fully exploiting the parallelism of the LRE algorithm with parallel Graphic Processing Unit (GPU) programming. Our result can play an important role in quantum information technologies with large quantum systems.

\end{abstract}

\pacs{03.67.-a,03.65.Wj}
\maketitle
\vspace{2pc}

\section{Introduction}
Quantum state tomography \cite{Pari04quantum,Haya05asymptotic,Lvov09continuous}, characterizing the state of a quantum system via quantum measurements and data processing, is a starting point and the standard for verification and benchmarking of various quantum information processing tasks, such as quantum computation \cite{Niel00quantum}, cryptography \cite{Gisi02quantum}, and metrology \cite{Giov04quantum, Giov06quantum, Giov11advances, Xian11entanglement, Crow12tradeoff}.

To reconstruct quantum states wherein we have no \emph{a priori} information, we can resort to informationally (over)complete measurements.  Quantum state tomography \cite{Liu04quantum, Liu05tomographic} using informationally (over)complete measurements is referred as \emph{full quantum state tomography} (FQST) in this paper. As there are ($d^2 -1$) independent parameters to characterize the density matrix of a $d$-dimensional quantum state, FQST needs at least ($d^2-1$) measurement operators. Note that the dimension $d$ grows exponentially with the size of the quantum system. Thus, the number of measurements and the computational complexity of data processing in FQST suffer the curse of dimensionality. Moreover, the time of data processing was found to be even much longer than the time required for implementing the measurements. As reported in \cite{Silv11practical,Smol12efficient,Haff05scalable}, reconstructing an 8-qubit state using maximum likelihood estimation (MLE) \cite{Pari04quantum,Jame01measurement} took almost a week, while the measurement time was only 10 hours \cite{Silv11practical}. With the breakthrough and rapid development of experimental techniques, the size of quantum systems with entanglement or coherence prepared in the laboratory has already grown from 2 qubits \cite{Kwia95new,Kwia99Ultrabright} to 10 qubits  in photonic systems (e.g., 8 qubits in \cite{Huan11experimental,Yao12observation} and 10 qubits in \cite{Gao10experimental}), 12 qubits in NMR \cite{NMR12qubit} and to even 14 qubits \cite{Monz1114qubit} in ion traps, overwhelming the capability of the available full quantum state tomography.

Much effort has been devoted to improving the performance of quantum state tomography \cite{PhysRevA.88.022101,PhysRevA.90.062123,PhysRevB.92.075312,Bartk16Piority,hou2016achieving}. Some methods focused on extracting partial concerned information. For example, entanglement witness \cite{Barb03detection,Bour04experimental} can detect entanglement with few measurements; direct purity estimation \cite{BagaBMR05} and fidelity estimation \cite{FlamL11,Silv11practical} were utilized to obtain the purity of the prepared state and its fidelity with the ideal state; permutationally invariant tomography \cite{Toth10permutationally} was used to extract information that will not change under permutation. Several approaches were concerned on performing quantum state tomography with \emph{a priori} knowledge or assumptions. For example, compressed sensing \cite{GrosLFB10,Weit12experimental,Kale15quantum} can perform quantum state tomography for quantum states with a low rank. If a quantum state is a matrix product state, it is possible to develop efficient tomography algorithms \cite{Baum13scalable,Cram10efficient}. However, these methods either extract partial information or have some prior information about the state to be reconstructed.

Several approaches have also been presented to reduce the computational complexity in the reconstruction algorithms in FQST \cite{Smol12efficient,Qi13quantum}. For example, the authors in \cite{Smol12efficient} developed an algorithm which can be used  to efficiently reconstruct a 9-qubit state in about five minutes. However, when the size of the quantum system increases one qubit, the running time will increase by a factor of more than ten according to Fig.~1 in \cite{Smol12efficient}, resulting in years of computation time for a 14-qubit state. In \cite{Qi13quantum}, a linear regression estimation (LRE) algorithm was proposed which has a much lower computational complexity than that of MLE for quantum state tomography \cite{Zhib15realization}.  In this paper, we push the data processing capability of FQST to a 14-qubit state using the informationally overcomplete Pauli measurements by optimizing the LRE algorithm in \cite{Qi13quantum} and employing parallel programming with graphics processing unit (GPU).

For experimental ease and high level of estimation accuracy, Pauli measurements are the preferred choice in experiments of FQST, although they are informationally overcomplete. LRE was demonstrated to be much more efficient than MLE in FQST \cite{Qi13quantum}. However, in order to reconstruct a 14-qubit state efficiently, we need to further optimize the LRE method. The efficiency of FQST in this paper refers solely to the reduced computational complexity of data processing. Our first optimization is  based on the fact that the representation of Pauli bases in the algorithm has very few nonzero elements under a proper choice of the representation. Furthermore, thanks to the simple LRE algorithm which only involves additions and multiplications of vectors and matrices, it is naturally suitable to be sped up by parallel programming. In parallel programming of matrix computation, GPU works much better than the central processing unit (CPU). Hence, we can use GPU parallel programming to realize the LRE algorithm and enhance the FQST capability. Compared with the result in \cite{Qi13quantum}, in this paper we optimize the LRE method with reduced computational complexity and storage requirement, and implement the optimized LRE method using GPU parallel programming. The results show significant enhancement of the FQST capability.

The rest of the paper is organized as follows. \Sref{sec: LRE} provides a brief introduction to the three steps of LRE and analyzes the computational complexity and the storage requirement in the case of informationally complete measurements. In \sref{sec: complexity}, computational complexity and storage requirement are discussed based on Pauli measurements. In \sref{sec: parallel GPU programming}, the LRE algorithm in the first two steps is realized using parallel GPU programming. The run time of the algorithm with GPU speeding up is also compared with that using CPU programming. In \sref{sec:accracy}, the estimation error of reconstructing a maximally-mixed state is analyzed in terms of squared Hilbert-Schmidt (HS) distance and infidelity. \Sref{sec:sum} presents the summary and prospect of this paper.

\section{\label{sec: LRE}Linear regression estimation algorithm}

In linear regression estimation (LRE) for a $d$-dimensional quantum system, the density matrix and measurement bases take a vector form after we choose a representation basis set $\{\Omega_{i}\}^{d^2-1}_{i=0}$ \cite{Qi13quantum}. The operators in this basis set are orthonormal, i.e.,  $\textmd{Tr}(\Omega^{\dag}_i\Omega_j)=\delta_{ij}$.  For convenience, let $\Omega_{i}=\Omega_{i}^{\dag}$ and all the bases are traceless except $\Omega_{0}=(1/d)^{\frac{1}{2}}I$. Elements in the vector form $\Theta$ of a quantum state $\rho$ are given by
\begin{equation}\label{eq:rho}
\theta_i=\textmd{Tr}(\rho\Omega_i).
\end{equation}
Given a set of $M$ measurement operators $\{|\Psi\rangle\langle\Psi|^{(j)}\}^{M}_{j=1}$, elements in the vector form $\Gamma^{(j)}$ of each $|\Psi\rangle\langle\Psi|^{(j)}$ are given by
\begin{equation}\label{eq:gamma}
\gamma^{(j)}_i=\textmd{Tr}(|\Psi\rangle\langle\Psi|^{(j)}\Omega_i).
\end{equation}

The whole LRE algorithm consists of three steps: \textbf{step (i)} Obtain the estimate of $\Theta$ using measurement data; \textbf{step (ii)} Construct a Hermitian matrix $\hat{\mu}$ satisfying $\textmd{Tr}\hat{\mu}=1$ from the estimate of $\Theta$; and \textbf{step (iii)} Find a physical density matrix $\hat{\rho}$ close to $\hat{\mu}$.

In step (i), a least-squared estimate $\hat{\Theta}^{LS}$, without consideration of the positivity restriction of quantum states, is given by
\begin{equation}\label{eq:theta}
\hat{\Theta}^{LS}=(X^{\top}X)^{-1}\sum^M_{j=1}\hat{p}_j\Gamma^{(j)},
\end{equation}
where $\hat{p}_j$ is the measured frequency of $|\Psi\rangle\langle\Psi|^{(j)}$, $X=\left(
        \begin{array}{ccc}
          \Gamma^{(1)}, & \cdots, & \Gamma^{(M)} \\
        \end{array}
      \right)^{\top}$ and $X^{\top}X=\sum_{j=1}^M \Gamma^{(j)}{\Gamma^{(j)}}^{\top}$.
The computational complexity in this step is at least $O(d^4)$, since FQST requires an informationally complete measurement set $\{|\Psi\rangle\langle\Psi|^{(j)}\}^{M}_{j=1}$ with $M=d^2-1$. For a 14-qubit state, $d=2^{14}$ and the computational complexity is $\sim 10^{16}$.

In step (ii), on the basis of the solution $\hat{\Theta}^{LS}$ to (\ref{eq:theta}) in step (i), we can obtain a Hermitian matrix $\hat{\mu}$ with $\textmd{Tr}\hat{\mu}=1$ by
\begin{equation}\label{eq:mu}
\hat{\mu}=\sum^{d^2-1}_{i=0}\hat{\theta}_i^{LS}\Omega_i.
\end{equation}
The computational complexity in this step is also $O(d^4)$.

The state estimate $\hat{\mu}$ obtained in \eqref{eq:mu} may have negative eigenvalues and be nonphysical due to the randomness of the finite measurement results. In step (iii), a proper method needs to be adopted to pull $\hat{\mu}$ back to a physical state. In this step, we use the fast algorithm proposed in \cite{Smol12efficient}, where a physical estimate $\hat{\rho}$ is chosen to be the closest density matrix to $\hat{\mu}$ under the matrix 2-norm. According to \cite{Smol12efficient}, the computational complexity in this step is $O(d^3)$.

It is clear that the computational complexity in LRE is dominated by the first two steps, which is $O(d^4)$. For a 14-qubit state, this is $\sim 10^{16}$. In terms of storage, $O(d^4)$ and $O(d^2)$ bytes are required  to store all the measurement bases and measurement results, respectively, for an informationally-complete measurement set. For a 14-qubit state, this needs tens of thousands of terabytes, which is beyond the capability for practical applications. In the following, we develop a method to reduce the computational complexity and storage requirement.

\section{\label{sec: complexity}Computational complexity and storage with Pauli measurements}

Pauli measurements are a good choice to extract information in quantum state tomography of $n$-qubit systems ($d=2^n$) because of not only their experimental ease but also the ability to achieve a high level of accuracy. With all the possible combinations of Pauli measurements for $n$-qubit systems, the total number of measurement bases is $M=6^n$. Without optimization, this informationally-overcomplete measurement set further increases the computational complexity in step (i) to $\sim 10^{19}$ for a 14-qubit state and also increases the storage requirement from $\sim 10^{16}$ to $\sim 10^{19}$.

The computational complexity and storage requirement can be greatly reduced because many terms in $\Gamma^{(j)}$ are zero when $\{\Omega_{i}\}^{d^2-1}_{i=0}$ are chosen as the tensor product of $\{\frac{1}{\sqrt{2}}\sigma_{i}\}^{3}_{i=0}$, with $\sigma_0=I_{2\times2}$, and Pauli matrices
$\sigma_1=\left(
                                                                \begin{array}{cc}
                                                                  0 & 1 \\
                                                                1 & 0 \\
                                                               \end{array}
                                                              \right)
$, $\sigma_2=\left(
                                                               \begin{array}{cc}
                                                                0 & -i \\
                                                                 i & 0 \\
                                                               \end{array}
                                                             \right)
$, and $\sigma_3=\left(
                                                                \begin{array}{cc}
                                                                  1 & 0 \\
                                                                  0 & -1 \\
                                                                \end{array}
                                                              \right)
$.  For an $n$-qubit state,
\begin{equation}\label{eq:Omega paoli}
  \Omega_{i}=2^{-n/2}\bigotimes\limits_{k=1}^n\sigma_{i_k}
\end{equation}
with $i_k=0,1,2,3,$ and $i=\sum\limits_{k=1}^ni_k\times4^{n-k}$.

In the 1-qubit scenario, inserting \eqref{eq:Omega paoli} into \eqref{eq:gamma}, one can easily obtain $\Gamma^{(j)}$ for all the eigenvectors of the three Pauli operators $\sigma_1,$ $\sigma_2,$ and $\sigma_3$, which are $\frac{1}{\sqrt{2}}(1,\pm1,0,0)^T$, $\frac{1}{\sqrt{2}}(1,0,\pm1,0)^T$, and $\frac{1}{\sqrt{2}}(1,0,0,\pm1)^T$. The term $X^{\top}X$ is calculated to be a diagonal matrix with $X^{\top}X=\text{diag}\{3,1,1,1\}$. As for the $n$-qubit scenario, since the measurement bases are the tensor product of those for single qubits, $\Gamma^{(j)}$ and $X^{\top}X$ are also the tensor products of their counterparts for relevant 1-qubit cases. Thus, there will be only $2^n$ nonzero elements in $\Gamma^{(j)}$, instead of the original $4^n$, and $X^{\top}X$ are diagonal. Dropping all zero elements will reduce the computational complexity in step (i) from $O(24^n)$ to $O(12^n)$, i.e., from $\sim 10^{19}$ to $\sim 10^{15}$, for a 14-qubit state.

After the optimization in step (i), the computational complexity $O(16^n)$ in step (ii) is dominant. Recall the definition of $\Omega_i$ in \eqref{eq:Omega paoli}, there are only $2^n$ nonzero elements. Therefore, after dropping all zero elements, the computational complexity in step (ii) can be reduced to $O(8^n)$ (i.e., $\sim 10^{12}$) for a 14-qubit state. Since the computational complexity in step (iii) is only $O(8^n)$, the total computational complexity with Pauli measurements is $O(12^n)$.

In terms of storage, without dropping the zero elements, $O(24^{n})$ bytes are needed to store all the measurement bases. Even after dropping all the zero elements, the storage requirement is still $O(12^{n})$ for the nonzero elements and their locations. However, the storage requirement of the measurement bases can be further reduced to $O(6^{n})$ after the $6^{n}$ bases are divided into $3^n$ groups. The $2^n$ operators in each group are all the eigenvectors of one combination of three Pauli operators. The $2^n$ nonzero elements in each of these $2^n$ operators in the same group share the same locations. All the nonzero values of operators in one group form a $2^n\times2^n$ nonzero matrix, which is the same for all the $3^n$ groups. Hence, we only need to store $3^n$ types of locations of nonzero elements (with $O(6^{n})$ storage requirement), and a $2^n\times2^n$ nonzero matrix (with $O(4^{n})$ storage requirement). Besides, the storage requirement for the measurement results of all the $6^n$ bases also requires $O(6^{n})$. Thus, all the needed storage is only $O(6^{n})$. Thus, the storage cost in the 14-qubit scenario is $\sim 10^{10}$, which is only tens of Giga Bytes.

\section{\label{sec: parallel GPU programming} Parallel GPU programming}

\begin{figure}
\center{\includegraphics[scale=0.9]{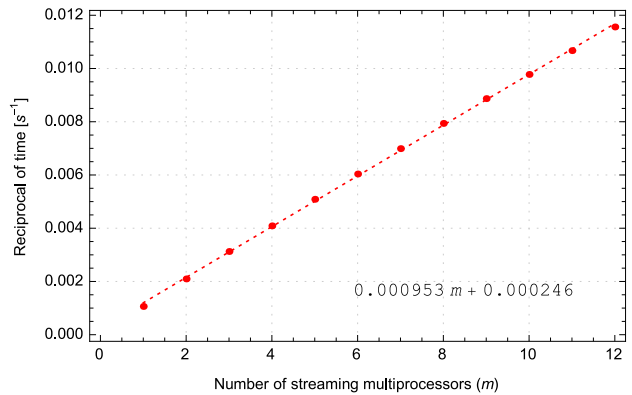}}
\caption{\label{fig:threads}
{Reciprocal of the run time versus the number of employed streaming multiprocessors (SMs). Red dots denote the reciprocal of the run time in step (i) for a 12-qubit state when $m$ SMs are used for parallel programming. Each SM has 192 CUDA cores. Since one thread is mapped to one CUDA core, $m$ SMs have 192$m$ threads running in parallel. The least-square linear fitting (red dotted line) shows that the speed of parallel programming increases almost proportionally with the number of parallel threads.}}
\end{figure}

\begin{figure*}[]
\centering
\subfigure[]{
\label{fig:first_step}
\includegraphics[width=0.43\textwidth]{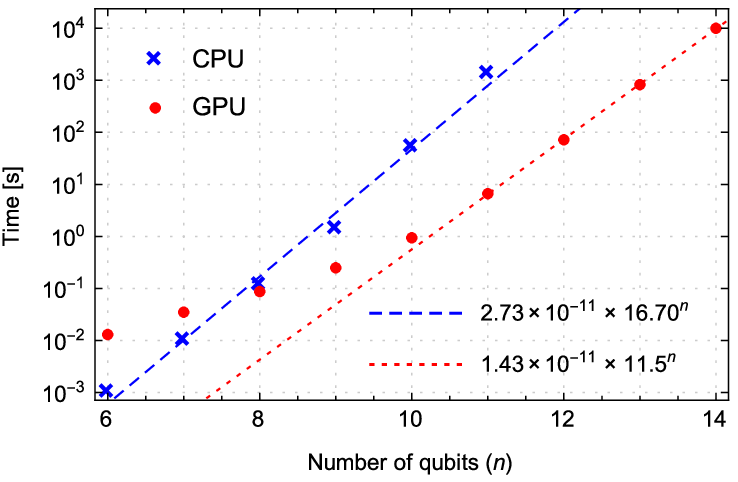}}
\subfigure[]{
\label{fig:second_step}
\includegraphics[width=0.43\textwidth]{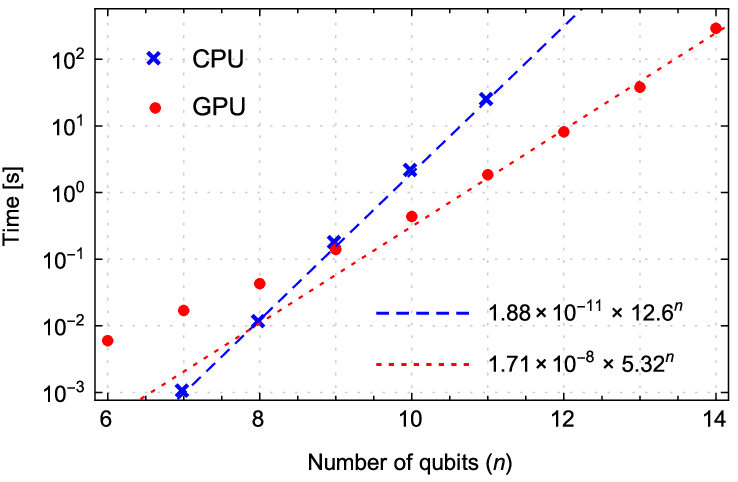}}
\subfigure[]{
\label{fig:third_step}
\includegraphics[width=0.43\textwidth]{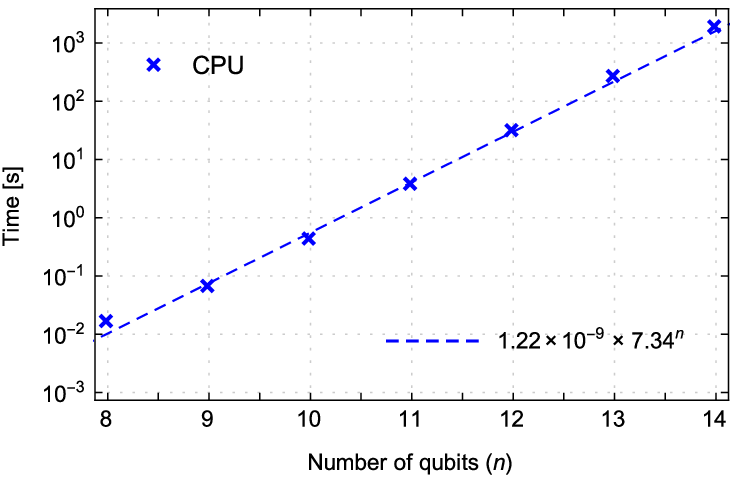}}
\subfigure[]{
\label{fig:total_step}
\includegraphics[width=0.43\textwidth]{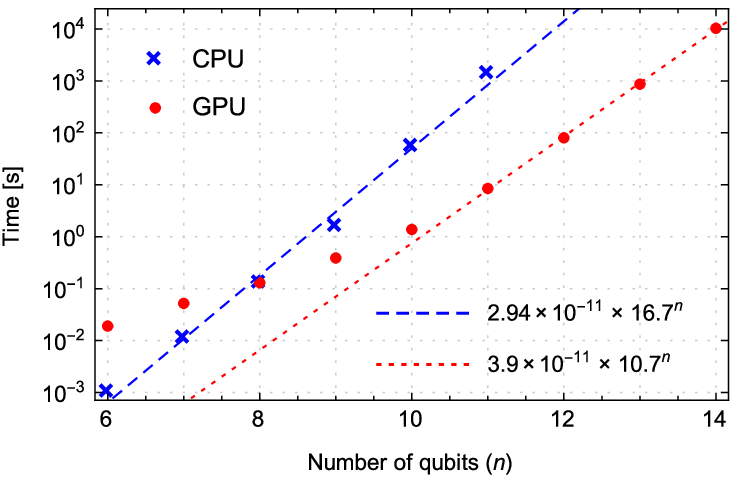}}
\caption{
(Color online) Computation time with respect to the number of qubits. Subfigures (a) to (d) depict, respectively, the computation time in steps (i)-(iii) and the whole process in the numerical reconstruction of a maximally-mixed state. Blue crosses and red dots represent the computation time for CPU and GPU programming of our optimized algorithm, respectively. When the number of qubits is larger than 8, GPU programming begins to show its advantage over CPU programming.  The dashed blue line is an exponential fitting of all the blue crosses. The dotted red line is an exponential fitting of the last four red dots because the time cost of GPU programming for a small sized system is dominated by the overhead of kernel calls.}
\end{figure*}

Thanks to the direct and simple formula of the LRE algorithm in \eqref{eq:theta} and \eqref{eq:mu}, only addition and multiplication operations of matrices or vectors are involved, which are naturally suitable for parallel programming. Graphic processing units possess powerful capability for parallel programming. This technique is exploited to speed up the computation in both step (i) and step (ii). The computer hardware includes 500 GB hard drive, 16 GB memory, i7-4770 CPU with 3.5 GHz, 4 cores and 8 threads, and GTX780 GPU with 2304 CUDA cores and 3G standard memory.

As analyzed in \sref{sec: complexity}, step (i) has the dominant computational complexity $O(12^n)$. GPU parallel programming is employed in this step. According to the locations of nonzero elements, all the measurement bases can be divided into $3^n$ groups. Each group has $2^n$ measurement bases. In parallel GPU programming, firstly, one group of measurement bases and their measurement data are put in, wherein the data include a $2^n\times2^n$ nonzero matrix, $2^n$ nonzero locations and $2^n$ measurement results. To be parallel, each thread is devoted to calculating one element of $\hat{\Theta}^{LS}$, whose location is in the $2^n$ nonzero locations. Hence, only $2^n$ elements of the total $4^n$ elements in $\hat{\Theta}^{LS}$ need to be calculated and updated for one group of data. All the threads are synchronized after updating these $2^n$ elements of $\hat{\Theta}^{LS}$ to get ready for the computation of the next group of data. Then this process continues until all the $3^n$ groups of data are computed.

{In order to show the relationship between the amount of speedup and the number of parallel threads in GPU programming, the reciprocal of the run time in step (i) for a maximally-mixed 12-qubit state is plotted with respect to $m$ in Fig.~\ref{fig:threads}, where $m$ is the number of streaming multiprocessors (SMs). Here, we only consider step (i) for simplicity, because step (i) has the dominant computational complexity. There are 12 SMs in the GTX780 GPU and each SM contains 192 CUDA cores. Since each CUDA core is occupied by a thread, $m$ SMs can execute 192$m$ threads in parallel. Fig.~\ref{fig:threads} shows that the speed in step (i), which is denoted by the reciprocal of the time cost, increases almost proportionally with the number of parallel threads, represented by the number of SMs. Hence, all the SMs are used in GPU programming to gain the fastest speed in the rest simulations.}

Using this approach we now demonstrate the computation time of reconstructing multi-qubit states using our algorithm. The computation time of GPU programming in step (i) costs 2.78 hours for a 14-qubit state as depicted in \fref{fig:first_step}. In comparison, to reconstruct a 11-qubit state, the CPU programming has taken almost half an hour for step (i) and will be difficult to compute larger systems as shown in \fref{fig:first_step}.

In step (ii), the time cost of CPU programming for a 14-qubit state approximates to be 10 hours according to the numerical fitting line (blue dashed line) in \fref{fig:second_step}, which is much longer than the run time of step (i) via GPU programming. Hence, GPU programming is also used to speed up the computation in this step. We note that the nonzero elements in $I_{2\times2}$ and $\sigma_z$ have the same locations, and so it is for $\sigma_x$ and $\sigma_y$. Thus all the $d^4$ $\Omega_{i}$, which are constructed in \eqref{eq:Omega paoli} by the above four matrices, can be divided into $2^n$ groups according to the same locations of nonzero elements. In parallel GPU programming, one group of $2^n$ $\Omega_{i}$ and the corresponding $2^n$ elements in $\hat{\Theta}^{LS}$ are put in. Each thread is responsible to calculate one of the corresponding $2^n$ elements in $\hat{\mu}$. All the threads are synchronized after finishing the update of the $2^n$ elements in $\hat{\mu}$ to get prepared for the next group of data. Then this step is repeated until all the $2^n$ groups of data are calculated. With parallel GPU programming, the computation time in step (ii) is reduced to only 0.08 hours, as shown in \fref{fig:second_step}.

\begin{figure*}
\center{\includegraphics[scale=1.3]{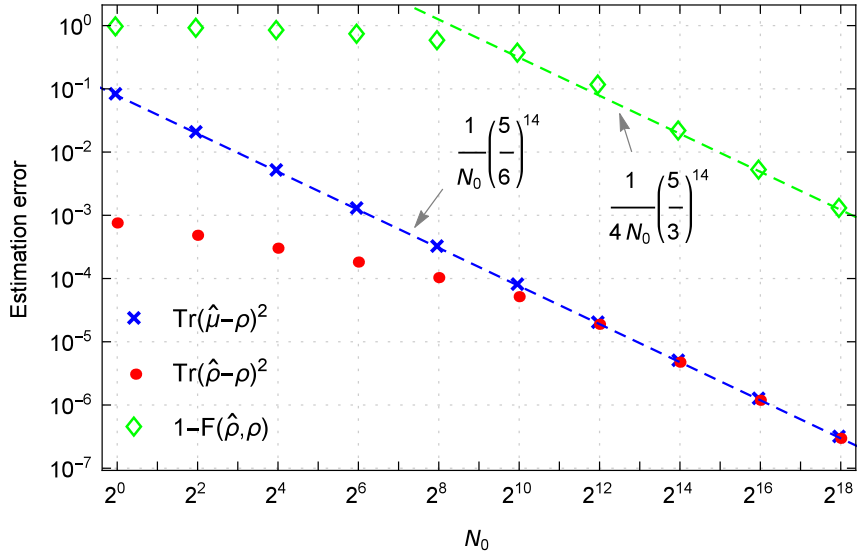}}
\caption{\label{fig:estimation error 14}
Estimation error of a maximally-mixed 14-qubit state with respect to different $N_0$. The total number of copies is $N=6^{14}N_0$. The numerical results of squared HS distance between $\hat{\mu}$ and $\rho$ (blue crosses) agree well with the asymptotic mean squared HS distances (dashed blue line). The squared HS distance between $\hat{\rho}$ and $\rho$ (red dots) is smaller than that between $\hat{\mu}$ and $\rho$, which is due to the positivity property \cite{Qi13quantum,Kale15quantum}. When the measured number of copies goes large, $\hat{\mu}$ is readily a physical state and will not change in step (iii). When $N_0$ gets larger than $2^{10}$, numerical results of infidelity between $\hat{\rho}$ and $\rho$ (green diamonds) match the dashed green line, which represents the infidelity between $\hat{\rho}$ and $\rho$ in the large number limit.}
\end{figure*}

In step (iii), we do not need GPU programming since the time cost of CPU programming with Mathematica is only 0.49 hours (see \fref{fig:third_step}) for a 14-qubit state, which is already about 5 times shorter than the computation time in step (i) using GPU programming. The total computation time of the whole process in reconstructing a 14-qubit state turns out to be only 3.35 hours, as shown in \fref{fig:total_step}. From \fref{fig:total_step}, we know that our optimized LRE algorithm based on CPU programming (blue) is already more than 100 times faster than the efficient algorithm in \cite{Smol12efficient} for reconstructing a 9-qubit state. However, according to the numerical fitting line in \fref{fig:total_step}, reconstructing a 14-qubit state will take more than a month using our optimized algorithm with CPU programming. This is also a reason why we resort to GPU programming for further speeding up through considering the parallelism of our algorithm. In our numerical simulations, all states are chosen as the maximally-mixed states in $n$-qubit systems only for the ease of obtaining the simulated measurement frequency. Our method is applicable to other states because the computational complexity and run time are state independent.

\section{\label{sec:accracy}Estimation error}

In this section, the error between the estimate and the real state is discussed in terms of the squared Hilbert-Schmidt (HS) distance and infidelity. The mean squared HS distance between the estimate state $\hat{\mu}$ in step (ii) and the true state $\rho$ is $\text{E}\textmd{Tr}(\hat{\mu}-\rho)^2=\text{E}(\hat{\Theta}^{LS}-\Theta)^{\top}(\hat{\Theta}^{LS}-\Theta)$, asymptotically given by
\begin{equation}\label{MSE}
\text{E}\textmd{Tr}(\hat{\mu}-\rho)^2=\frac{M}{Nd}\textmd{Tr}[(X^{\top}X)^{-1}X^{\top}PX(X^{\top}X)^{-1}],
\end{equation}
where $N$ is the total number of copies for all the $M$ operators and $P=\text{diag}(p_1-p_1^2,\cdots,p_M-p_M^2)$, with $p_j=\textmd{Tr}(\rho|\Psi\rangle\langle\Psi|^{(j)})$. Here we have one more factor of $1/d$ than Eq.~(9) in \cite{Qi13quantum} since every $d$ measurement bases form a complete set of POVM and can be measured simultaneously. As these $d$ measurement bases are measured simultaneously, the corresponding off-diagonal elements in $P$ are of $O(p_ip_j)$ rather than zero. But these off-diagonal elements are much smaller than the diagonal elements when the size of the quantum system is large. Therefore, $P$ is still approximated to only have diagonal elements in the case of simultaneous measurements.

This mean squared HS distance obviously depends on the unknown state. Here we calculate the mean squared HS distance when $\rho$ is chosen as the maximally-mixed state, i.e., $\rho=\frac{I_{d\times d}}{d}$. One reason for this state choice is that an elegant theoretical result can be derived due to the simple form of a maximally-mixed state; the other reason is that the maximally-mixed state often gives the larger mean squared HS distance than the other states. In this case, we have $P=\frac{I_{d\times d}}{d}$ in the first order of approximation. When Pauli measurements are performed on copies of a $n$-qubit maximally-mixed state, the mean squared HS distance (blue line in \fref{fig:estimation error 14}) is equal to $\frac{1}{N_0}(\frac{5}{6})^n$ with $N_0=\frac{N}{M}$. This theoretical result agrees well with the numerical results of $\textmd{Tr}(\hat{\mu}-\rho)^2$ (blue crosses in \fref{fig:estimation error 14}). Thanks to the positivity condition, compared with $\textmd{Tr}(\hat{\mu}-\rho)^2$, the squared HS distance of a physical estimate $\hat{\rho}$ for the real state (denoted as red dots in \fref{fig:estimation error 14}) is further reduced for $N_0\leq2^{10}$ although it remains the same when $N_0>2^{10}$ for a 14-qubit maximally-mixed state.

Another well-motivated figure of merit is infidelity, defined as $1-F(\hat{\rho},\rho)=1-\mathrm{Tr}^2(\sqrt{\sqrt{\rho}\hat{\rho}\sqrt{\rho}})$. When $\rho$ is chosen as the maximally-mixed state, infidelity is directly related to the squared HS distance by $1-F(\hat{\rho},\rho)=\frac{d}{4}\textmd{Tr}(\hat{\rho}-\rho)^2+d^2O(\textmd{Tr}(\hat{\rho}-\rho)^3)$. In the large-number limit of $N_0$, $\hat{\mu}$ has non-negative eigenvalues and we have $\hat{\mu}=\hat{\rho}$. Thus, the average infidelity is well approximated by $\frac{1}{4N_0}(\frac{5}{3})^n$ (green line) in this limit. However, when $N_0$ is small, $\hat{\mu}$ will have negative eigenvalues and needs to be pulled back to a physical state $\hat{\rho}$. Hence, as shown in \fref{fig:estimation error 14}, the real infidelity (green diamonds) is much smaller than the prediction (green line) for $N<2^{10}$ in the 14-qubit scenario.

\section{\label{sec:sum}Summary and prospect}
In this paper, we have fully reconstructed a 14-qubit state with a modest computation time of 3.35 hours using informationally overcomplete Pauli measurements. By a smart choice of the representation basis set in the algorithm, only very few nonzero elements exist and need to be stored and calculated, which reduce the computational complexity by a factor of $2^n$. Furthermore, parallel GPU programming is used to fully exploit the parallelism in the simple LRE algorithm. It is worth pointing out that our method can push the capability of FQST forward to even larger quantum systems. This is because the measurement bases in our analysis and simulations are chosen as Pauli measurements, which are overcomplete. If we reduce the number of measurement bases from $6^n$ to $4^n$, the computational complexity can be reduced to $O(8^n)$ from $O(12^n)$, which can further enhance the capability of FQST via our optimized LRE.

\section*{Acknowledgments}

Z. H., H. Z. and Y. T. thank Bin Zhou for his lectures in GPU programming. GTX 780 used for this research was donated by the NVIDIA Corporation. The authors would like to thank Adam Miranowicz for helpful suggestions that have improved this paper. The work was supported by the National Natural Science Foundation of China under Grants (Nos.  61222504, 11574291, 61374092 and 61227902) and the Australian Research Council's Discovery Projects funding scheme under Project DP130101658. FN is partially supported by the RIKEN iTHES Project, the IMPACT program of JST, and a Grant-in-Aid for Scientific Research (A).

%
\providecommand{\newblock}{}

\end{document}